\documentclass[usenatbib]{mn2e}

\usepackage{graphicx}
\usepackage{amssymb}
\usepackage{amsmath}
\usepackage{color}

\newcommand{\be}{\begin{equation}}
\newcommand{\ee}{\end{equation}}

\newcommand{\fracp}[2]{\left(\frac{#1}{#2}\right)}

\newcommand{\beq}{\begin{equation}}
\newcommand{\eeq}{\end{equation}} 
\newcommand{\ind}[1]{{\rm{#1}}}

\begin{document}

\title[ Solution to the Sigma-Problem ]
{Solution to the Sigma-Problem of Pulsar Wind Nebulae.}

\author[Porth et al.]{Oliver Porth$^{1,2}$\thanks{E-mail: o.porth@leeds.ac.uk (OP)}, 
Serguei S. Komissarov$^{1}$\thanks{E-mail: serguei@maths.leeds.ac.uk (SSK)}, 
Rony Keppens$^{2}$\\
$^{1}$Department of Applied Mathematics, The University of Leeds, Leeds, LS2 9GT \\
$^{2}$Centre for mathematical Plasma Astrophysics, Department of Mathematics, KU Leuven,
 Celestijnenlaan 200B, 3001 Heverlee, Belgium}

\date{Received/Accepted}
\maketitle

\begin{abstract} 
We present first results of  three dimensional relativistic 
magnetohydrodynamical simulations of Pulsar Wind Nebulae. They show 
that the kink instability and magnetic dissipation inside these nebulae 
may be the key processes allowing to reconcile their observations with 
the theory of pulsar winds. In particular, the size of the termination 
shock, obtained in the simulations, agrees very well with the observations 
even for Poynting-dominated pulsar winds. Due to magnetic dissipation 
the total pressure in the simulated nebulae is particle-dominated and more or 
less uniform.  While in the main body of the simulated nebulae the magnetic 
field becomes rather randomized, close to the termination shock, it is dominated 
by the regular toroidal field freshly-injected by the pulsar wind. This field is responsible 
for driving polar outflows and may explain the high polarization observed in 
pulsar wind nebulae.         
\end{abstract}

\begin{keywords}
ISM: supernova remnants -- MHD -- instabilities -- relativity -- shock waves -- pulsars: general -- pulsars: individual: Crab
\end{keywords}

\section{Introduction}

It is now well established that Pulsar Wind Nebulae (PWN) are powered
by relativistic winds from neutron stars, formed during violent deaths
of massive normal stars. Theory of such winds, based on the physical
processes in the magnetospheres of neutron stars, predicts that they
are strongly Poynting-dominated at their base \citep[see ][and
  references therein]{arons12}.

However, simple one-dimensional models of PWN fit the observations
only if pulsar winds are particle-dominated.  Quantitatively this
conflict is best described using the wind magnetization parameter
$\sigma$ defined as the ratio of the wind Poynting flux to its kinetic
energy flux. Theoretical models of pulsar magnetospheres and wind
predict $\sigma\gg1$, whereas the 1D models of PWN suggest
$\sigma\sim10^{-3}$ \citep{rees-gunn-74,kc84a,emm-che-87,begelman-92}.
A slightly higher magnetization, $\sigma\sim10^{-2}$, was later
suggested by axisymmetric numerical simulations
\citep{ssk-lyub-03,ssk-lyub-04,delzanna-04,bogovalov-05}.

Several possible explanations of this $\sigma$-problem have been put
forward over the years. The simplest one is that the electromagnetic
energy of the pulsar wind is converted into kinetic energy of the wind
on its way from the pulsar to the wind termination shock, where
freshly supplied plasma is injected into the PWN. Although claims
have been made that ideal MHD acceleration mechanisms can provide the
required energy conversion \citep[e.g.][]{Vlahakis-04}, it has now
become clear that this is not the case
\citep[e.g.][]{kvkb-09,lyub-09,lyub-10}.

The acceleration can be facilitated via non-ideal processes, involving
magnetic dissipation, in pulsar winds \citep{coroniti-90}.  For the wind
of the Crab pulsar, the dissipation length scale still significantly
exceeds the radius of the wind termination shock \citep{lyub-kirk-01},
unless the pulsar produces much more plasma compared to the
predictions of the current models of pair-production in pulsar
magnetospheres \citep[see ][and references therein]{arons12}.
Alternatively, the energy associated with the alternating component of
magnetic field of the striped wind can be rapidly dissipated at the
termination shock itself \citep{lyub-03,sironi-spt-11}.
Although the striped-wind model allows conversion of a large fraction
of the total Poynting flux into the internal energy of PWN plasma,
much more is needed to approach the target value of
$\sigma\sim10^{-3} \div 10^{-2}$. Indeed, the dissipation is confined to the
striped zone of the wind and only the alternating component of
magnetic field dissipates. Closer to the poles, the magnetization of
pulsar wind plasma remains very high even after it crosses the
termination shock and, as a result, the overall magnetization of
plasma injected into the nebula is much higher than that of the
Kennel-Coroniti model, unless the striped zone spreads over almost the
entire wind, implying that the pulsar is an almost orthogonal rotator
\citep{coroniti-90,ssk-mdiss-12}.

In contrast to these ideas, \citet{begelman-98} argued that the 1D
models of PWN could be highly unrealistic and their predictions not
trustworthy.  He has shown that the plasma configuration assumed in
these models is unstable to the magnetic kink instability and
speculated that the disrupted configuration may be less demanding on
the magnetization of pulsar winds. Indeed, one would expect the
magnetic pressure due to randomized magnetic field to dominate the
mean Maxwell stress tensor, and the adiabatic compression to have the
same effect on the magnetic pressure as on the thermodynamic pressure
of relativistic gas. Under such conditions, the global dynamics of PWN produced
by high-$\sigma$ winds may not be that much different from those of PWN
produced by particle-dominated winds.  These expectations have received
strong support from the recent numerical studies \citep{mizuno-11} of
magnetic kink-instability of the cylindrical magnetostatic
configuration, which was used in \citet{begelman-92} to model
PWN. These simulations have shown a relaxation towards quasi-uniform
total pressure distribution inside the computational domain on the
dynamical time-scale.  \citet{mizuno-11} also reported a significant
magnetic dissipation.  

The most important factor missing in the simulations by
\citet{mizuno-11} is the continuous injection of magnetic flux and
energy in PWN by their pulsar winds. As a result, there is no
termination shock whose size is an important parameter used to test
theories of PWN against the observational data. The observed strong
polarization of PWN is suggestive of significant regular magnetic
field in the form of concentric loops in their central regions, which
could have been supplied by the pulsar wind relatively recently. The
polar jets, most prominent in the Crab and Vela nebulae are best
explained by the action of magnetic hoop stress of such fields
\citep{lyub-02,ssk-lyub-03,delzanna-04, bogovalov-05}.  Thus, the next
logical step is to carry out 3D numerical simulations of PWN proper,
similar in their setup to the previous axisymmetric simulations. We
are currently conducting such studies and here report its first most interesting results.

\section{Simulation setup}

Initially, the computational domain is split in two zones separated by
a spherical boundary of radius $r_i=10^{18}$cm. The outer zone is
filled with radially-expanding cold supernova ejecta. The
solution in this zone has Hubble's velocity profile 
\beq 
v=v_\ind{i}
\fracp{r}{r_\ind{i}} 
\eeq 

and constant density $\rho_\ind{e}$. The values of parameters 
$\rho_\ind{e}$ and $v_\ind{i}$ are determined using the observed 
expansion rate of the Crab nebula and typical parameters of type-II 
supernovae. 
The inner zone is initially filled with unshocked pulsar wind. In our
model of the wind, we assume that the alternating component of magnetic
field in its striped zone has completely dissipated along the way from
the pulsar to the PWN. Although this may not be the case and most of
the dissipation occurs instead at the termination shock, dynamically
this makes no difference \citep{lyub-03}. The total energy flux
density of the wind follows that of the monopole model
\citep{michel-73}

\begin{align}
f_{\rm tot} (r,\theta) \propto \left(\frac{1}{r^{2}}\right)
\left(\sin^{2}\theta+b\right) \, .
\label{eq:ftot}
\end{align}
The parameter $b=0.03$ is added for numerical reasons.  This
energy is divided into magnetic, $f_{\rm m}$,  and kinetic, $f_{\rm k}$, 
terms

\begin{align}
f_{\rm m}(r,\theta) = \sigma(\theta)\frac{f_{\rm
    tot}(\theta,r)}{1+\sigma(\theta)}, \quad
f_{\rm k}(r,\theta) = \frac{f_{\rm tot}(r,\theta)}{1+\sigma(\theta)} \,,
\label{eq:fs}
\end{align}
where $\sigma(\theta)$ is the latitude dependent wind magnetization.

\begin{figure}
\begin{center}
\includegraphics[width=80mm]{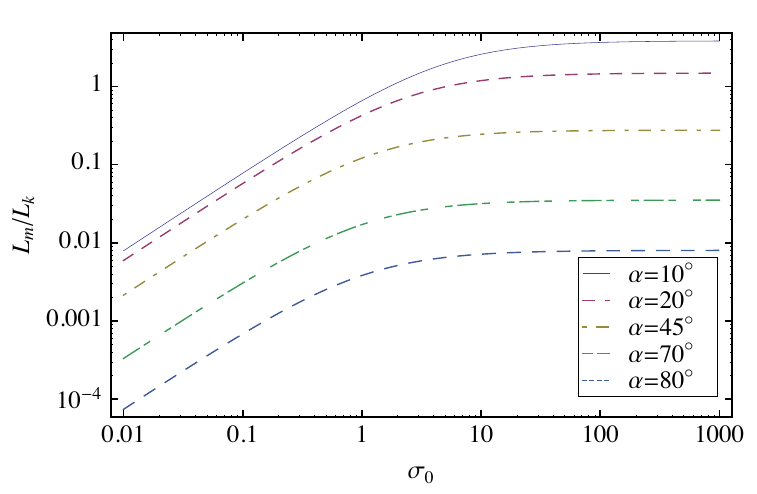}
\caption{Electromagnetic luminosity over kinetic luminosity injected
  into the PWN against value before reconnection $\sigma_{0}$ for
  various angles of obliqueness $\alpha$.  If the obliqueness is larger than $\sim
  45^{\circ}$, the luminosity fraction saturates at approximately this
  value.  This is why our dynamical simulations can also provide
  reasonable models for highly magnetized cases with $\sigma_{0}>
  10^{3}$.  }
\label{fig:lmOverLk}
\end{center}
\end{figure}

\begin{figure*}
\begin{center}
\includegraphics[width=75mm]{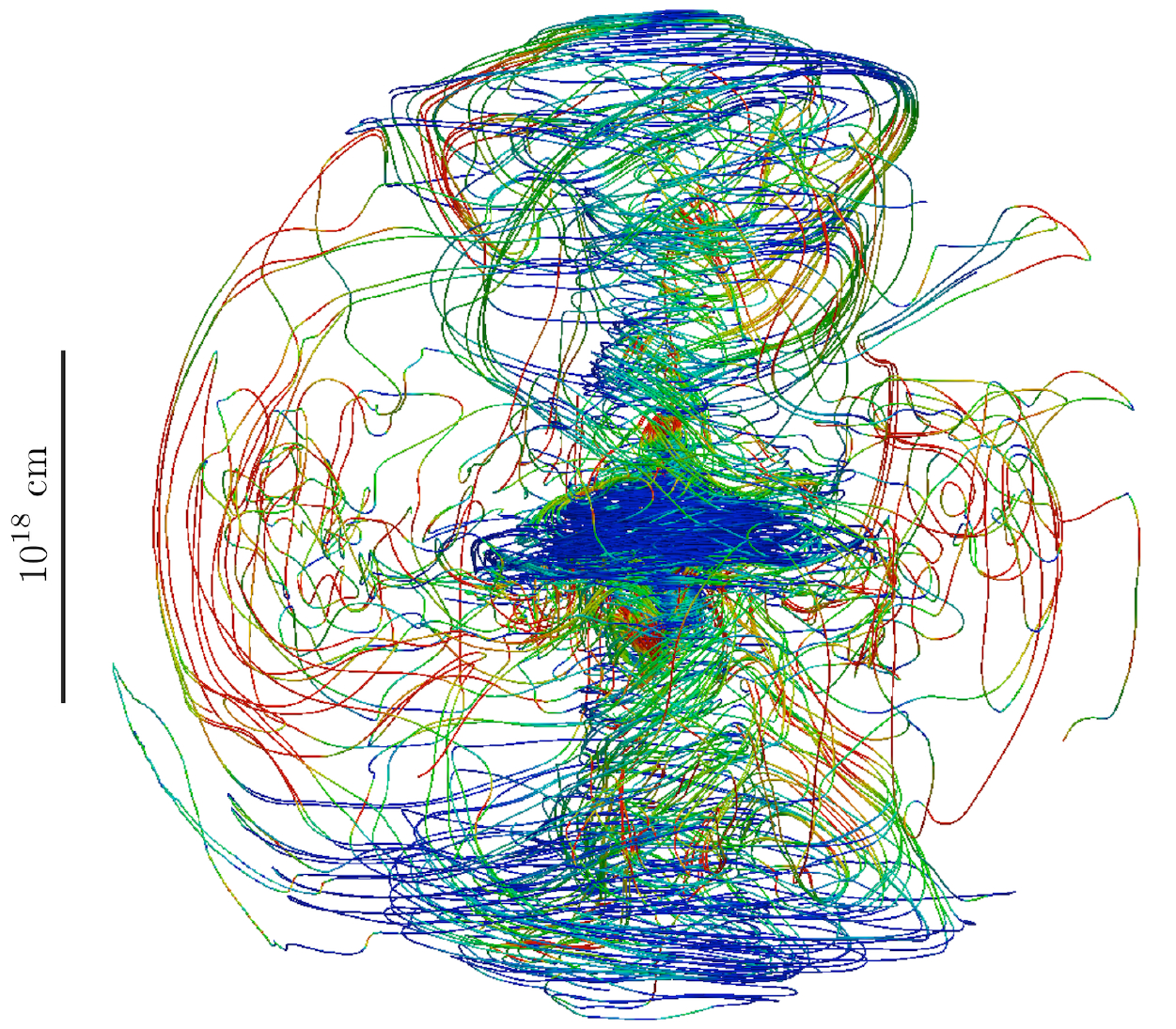}
\includegraphics[width=50mm]{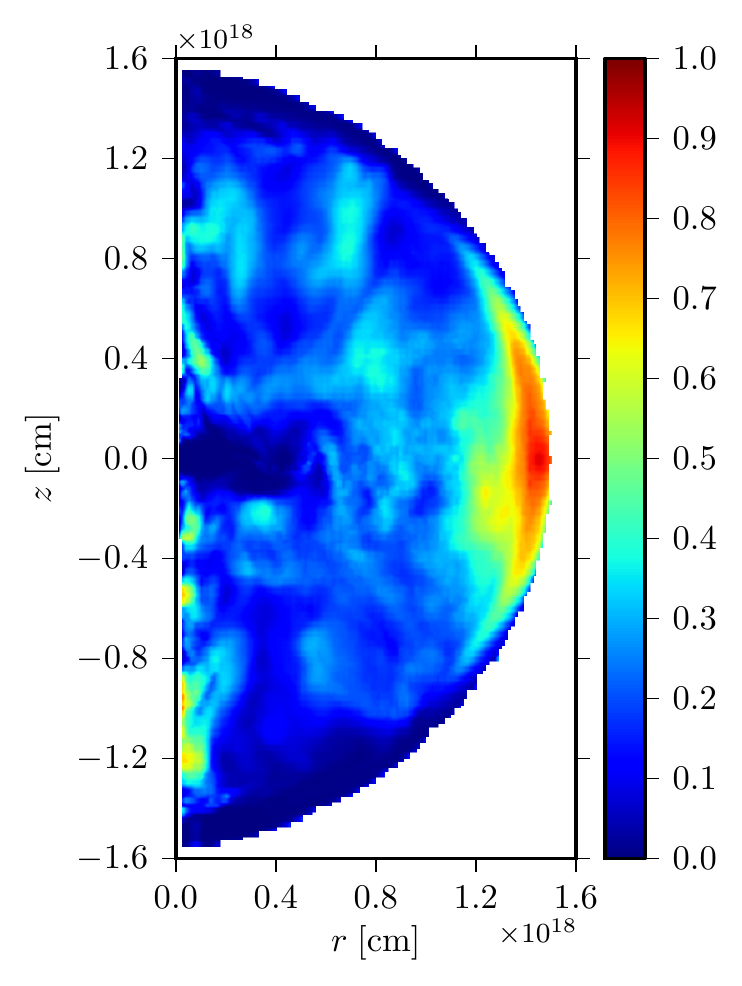}
\caption{Left panel: 3D rendering of the magnetic field structure in 
the model with $\sigma_{0}=3$ at $t\simeq70\rm~years$ after the start of the simulation.  
Magnetic field lines are integrated from sample points
starting at $r=3\times 10^{17}\rm cm$. Colors indicate the
dominating field component, blue for toroidal and red for poloidal. 
Right panel: Azimuthally averaged 
$\bar{\alpha}\equiv\langle B_{p}^{2}/B^{2}\rangle_{\phi}$ for
the same model.  
}
\label{fig:3drender}
\end{center}
\end{figure*}

For simplicity, we assume that before dissipation of magnetic stripes
the wind magnetization saturates at
\begin{align}
\tilde{\sigma}_{0}(\theta) = \left\{
\begin{array}{ll}
\left(\theta/\theta_{\rm 0}\right)^{2}\ \sigma_{0}; &
\theta\le\theta_{\rm 0}\\ \sigma_{0}; & \theta > \theta_{\rm 0}
\end{array}
\right.
\label{eq:sig0}
\end{align}
where the additional (small) parameter $\theta_{0}=10^{\circ}$ was introduced
to ensure that the Poynting flux vanishes on the axis for
an axisymmetric model.  Dissipation of magnetic stripes changes the
magnetization of the striped wind zone so that

\begin{align}
\sigma(\theta) =
\frac{\tilde{\sigma}_{0}(\theta)\chi_{\alpha}(\theta)}{1+\tilde{\sigma}_{0}(\theta)(1-\chi_{\alpha}(\theta))}\,
,
\end{align}
where 

\beq \chi_\alpha(\theta) = \left\{
    \begin{array}{ll}
       (2\phi_\alpha(\theta)/\pi-1)^2, & \pi/2-\alpha<\theta<\pi/2
      +\alpha \\ 1, & \mbox{otherwise}
\end{array}
\right. , 
\eeq 
with $\phi_{\alpha} (\theta)\equiv \arccos(-\cot(\theta)\cot(\alpha))$.  Here, $\alpha$ signifies the magnetic inclination angle of the
pulsar which determines the size of the striped wind zone
\citep{ssk-mdiss-12}. Figure~\ref{fig:lmOverLk} shows the ratio of the
wind total electromagnetic luminosity to its total kinetic luminosity, 
obtained via integration of the fluxes given by Eqs.~\ref{eq:fs},
as a function of $\sigma_0$ for different values of the magnetic 
inclination angle. In this letter we present simulations with 
$\sigma_0=0.01,1,3$ and $\alpha=45^o$. As one can see in Fig.~\ref{fig:lmOverLk}, 
for $\sigma_0=3$ the luminosity ratio is already very close to the 
asymptotic value in the limit $\sigma_0\to \infty$, which is determined by the 
extent of the striped wind zone for all except very small values of $\alpha$.      
This is why we expect our models with $\sigma_0=3$ not to be very different 
from those with much higher $\sigma_0$, which can be found in real 
pulsar winds.

The wind's magnetic field is purely azimuthal and changes direction at the 
equatorial plane. Its strength, as measured in the pulsar frame, is found via
\beq 
B_{\phi}(r,\theta) = \pm\sqrt{4\pi f_{\rm m} (r,\theta)/v}\, , 
\eeq
where $v$ is the radial wind velocity. For simplicity, we adopt the same value
of the wind Lorentz factor, $\Gamma=10$, for all its streamlines.  
Hence the numerical wind density follows as 
\beq
\rho(r,\theta) = f_{\rm k}(r,\theta) / (\Gamma^{2}c^{2}v).
\eeq

When the simulations begin, the discontinuity at $r_\ind{i}$ between these two
initial zones splits and the termination shock first rapidly
moves inside towards the wind origin, reflecting the artificial nature
of our initial configuration.  Soon after, it re-bounces and begins to
expand at a much slower rate, gradually approaching its asymptotic
self-similar position \citep{rees-gunn-74}.  For an unmagnetized wind,
$\sigma_0=0$, its equatorial radius at this stage evolves as
 
\begin{align}
r_{0} \simeq \sqrt{2} r_{\rm n} \left(\frac{v_{\rm n}}{c}\right)^{1/2}
\left( 1 - \left( 1+ \frac{v_\ind{n}t}{r_\ind{i}} \right)^{-2}
\right)^{-1/2}\, ,
\label{eq:r0hydro}
\end{align}
where $v_\ind{n}$ is the expansion rate of the PWN, and $t$ is the
time since the start of the simulations.  One can see that the
time-scale of transition to the self-similar expansion is $\sim
r_\ind{i}/v_\ind{i}$, which is $\simeq210~$yr for the parameters of 
our simulations. We do not yet fully reach the self-similar 
phase in 3D simulations and hence use this solution as a reference.

The simulations are performed with MPI-AMRVAC \citep{amrvac}, solving
the set of special relativistic magneto-hydrodynamic conservation laws
in cartesian geometry.  We employ a cubic domain with edge length of
$2\times 10^{19}\rm cm$, large enough to contain today's Crab
nebula. Outflow boundary conditions allow plasma to leave the
simulation box.  The adaptive mesh refinement starts at a base level
of $64^{3}$ cells and is used to resolve the expanding nebula bubble
with a cell-size of $\Delta x=1.95\times 10^{16} \rm cm$ (on the fifth
level).  
Special action is taken to properly resolve the termination shock and
the flow near the origin.  To this end, additional grid levels
centered on the termination shock are automatically activated,
depending on the shock size. For the simulations shown here, the shock
is thus resolved by $3-4$ extra grid-levels (resulting in $8-9$ levels in total) 
on which we employ a more
robust minmod reconstruction in combination with Lax-Friedrich flux
splitting.  2D comparison simulations are performed with equivalent
numerical setup and differ only in the use of cylindrical coordinates
in the $r,z$-plane.

\begin{figure*}
\begin{center}
\includegraphics[width=125mm]{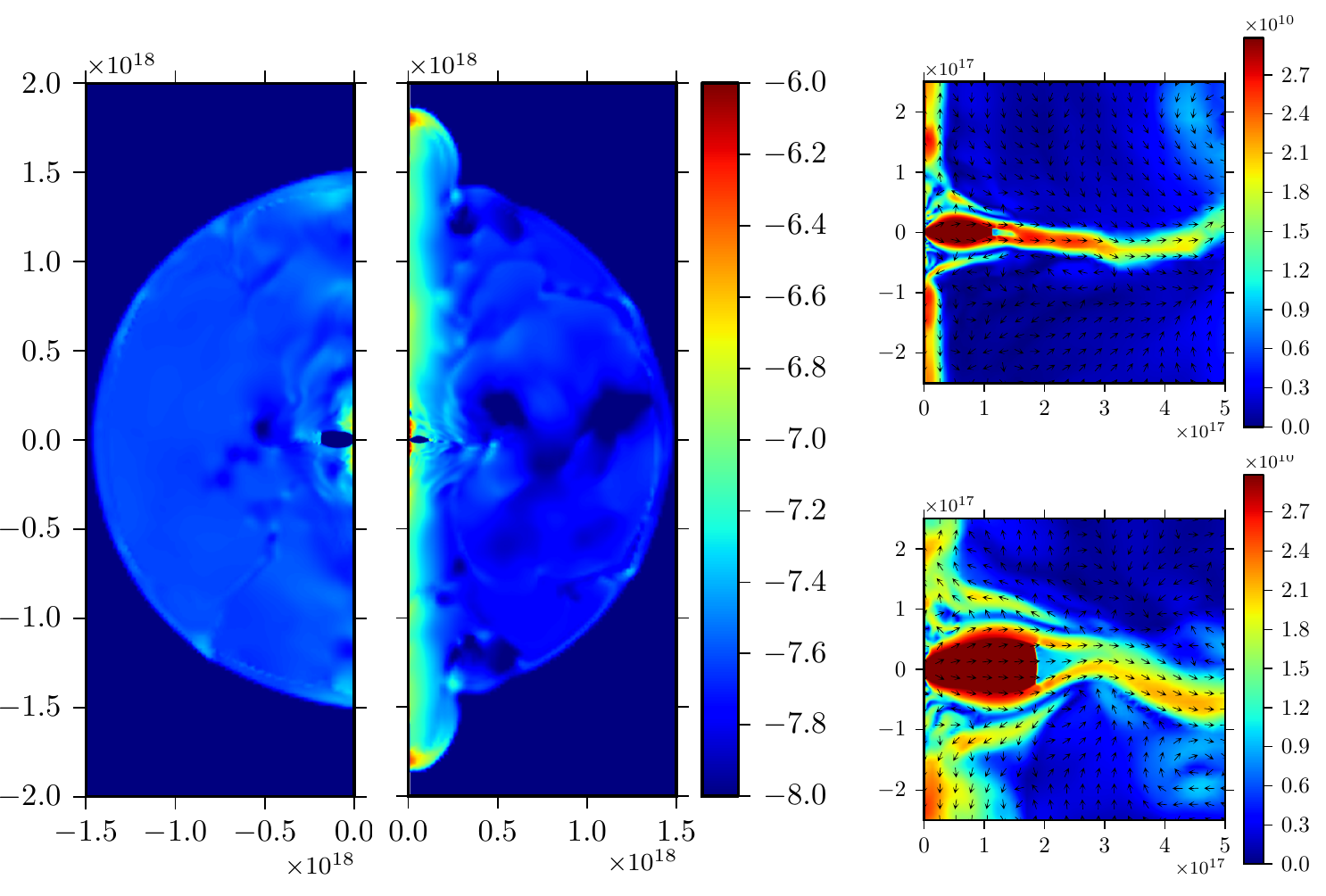}
\caption{Left panel: The total pressure distribution in the 3D 
(left hemisphere) and the corresponding 2D (right hemisphere) solutions 
for the same model with $\sigma_0=3$ at $t\simeq70 \rm~years$.  
Right panels: Velocity magnitude and direction for the same solutions, 
with the 2D one at the top and 3D one at the bottom. Note the scale difference between left/right panels. }
\label{fig:2Dvs3D}
\end{center}
\end{figure*}

\section{Results}

In basic agreement with the simulations by \citet{mizuno-11}, the highly 
organized coaxial configuration of magnetic field, characteristic of 
previous 2D simulations of PWN, is largely destroyed in our 3D models.                
However, the azimuthal component is still dominant in the vicinity of 
the termination shock, in the region roughly corresponding to Crab's torus 
(see Figure~\ref{fig:3drender}), which is filled mainly with ``fresh'' plasma 
which is just on its way from the termination shock to the main body of 
the nebula. As we have pointed out in the introduction, the emission of 
this plasma could be behind the strong polarization observed in the central 
region of the Crab nebula.  This figure also shows predominantly poloidal 
magnetic field in the outskirts of PWN, close to the equator. However, 
the magnetic field is rather weak in this region.  

Also in agreement with \citet{mizuno-11}, the total pressure distribution of 
our 3D solutions is much more uniform compared to 2D solutions of the same 
problem. As a result, the expansion of PWN in 3D is more or less isotropic, 
whereas in 2D the artificially enhanced axial compression due to the magnetic 
hoop stress promotes noticeably faster expansion in the polar direction. 
As one can see in Figure~\ref{fig:2Dvs3D}, by the end of the 
simulations, the 2D solution with $\sigma_0=3$ begins to exhibit a jet breakout, 
similar to those observed in the earlier 2D simulations of 
highly-magnetized young PWN of magnetars \citep{bucc-07,bucc-08}, with 
application to Gamma Ray Bursts.                 

\begin{figure}
\begin{center}
\includegraphics[width=80mm]{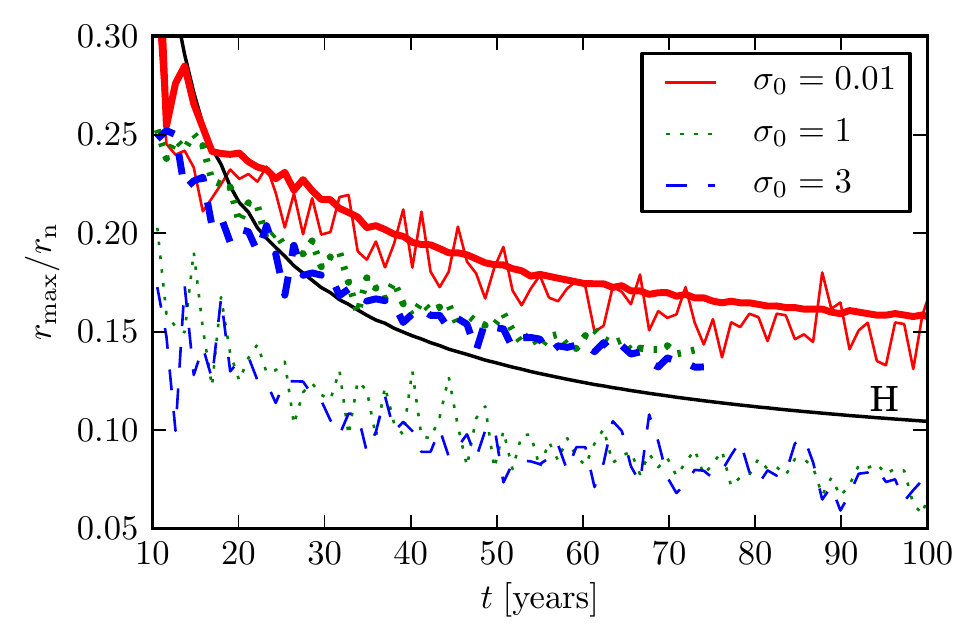}
\caption{Shock size in the equatorial plane for the two-
  \textit{(thin lines)} and three-dimensional case \textit{(thick lines)}.  The
  solid black curve labeled ``H'' follows the hydrodynamic prediction according
  to Eq.~\ref{eq:r0hydro}.}
\label{fig:rshock}
\end{center}
\end{figure}

Figure~\ref{fig:rshock} shows the evolution of the equatorial radius of the 
termination shock with time in all our simulations together with the 
analytical prediction from Eq.~\ref{eq:r0hydro}.  We anticipated that in 3D the 
shock radius turns out similar to that given by the analytical 
model for unmagnetized wind, as this was suggested in \citet{begelman-98}. 
This is indeed the case, but Figure~\ref{fig:rshock} 
also shows that the size of the termination shock exhibits 
only a weak dependence on the initial magnetization $\sigma_{0}$, once 
$\sigma_{0}\ge1$. However, we also expected to find a much smaller 
shock radius in 2D simulations, as their symmetry prevents the development of 
the key process in the Begelman's theory, the kink instability. What we have 
actually found is that, although in our 2D simulations the shock radius is indeed 
smaller, the difference is not dramatic.  

\begin{figure}
\begin{center}
\includegraphics[width=80mm]{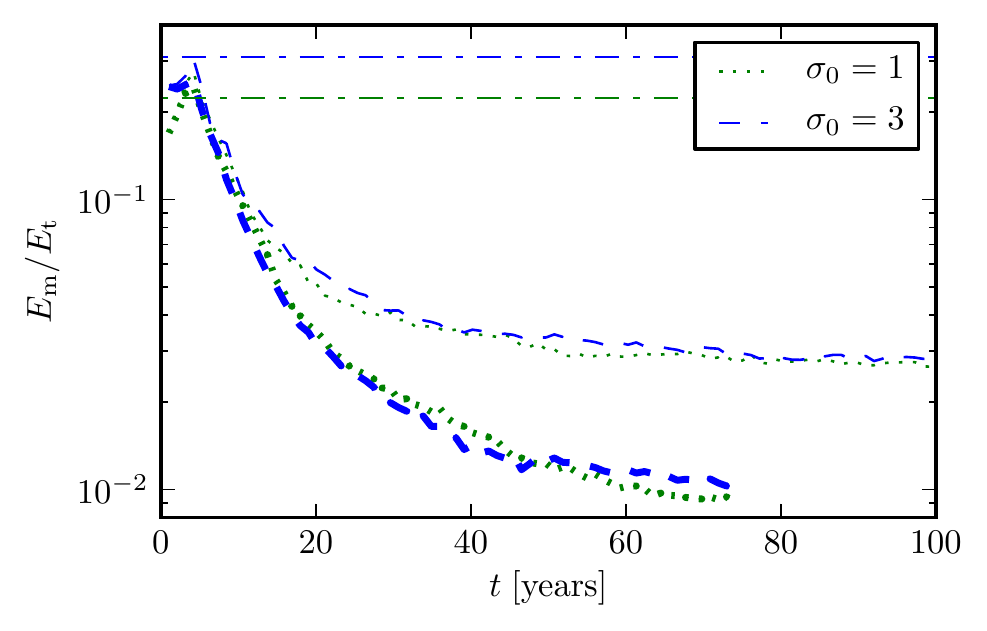}
\caption{Ratio of the total magnetic energy to the total thermal energy 
of PWN as a function of time for 2D \textit{(thin lines)} and 3D 
\textit{(thick lines)} models. Horizontal dash-dotted lines indicate the 
expected ratios at the self-similar phase in the absence of magnetic 
dissipation \citep{ssk-mdiss-12}.} 
\label{fig:mdiss}
\end{center}
\end{figure}

An explanation for this result is suggested by 
Figure~\ref{fig:mdiss}, which shows the ratio of magnetic to 
thermal energy of simulated PWN in our numerical models. 
One can see that, not only do our 3D solutions exhibit significant magnetic 
dissipation, which agrees with \citet{mizuno-11}, but the 2D solutions do 
as well. Indeed, for 2D models with $\sigma_0=1,3$ 
this parameter  decreases from $\sim 0.3$, which is close to the mean value 
of freshly-injected plasma \citep{ssk-mdiss-12}, to $\sim 0.03$. 
Such a low value means that the total pressure in the nebula is dominated 
by the gas pressure, just like in models with particle-dominated winds, and 
this is why the shock radius is close to that of the unmagnetized model. 
The property of our 2D models which makes magnetic dissipation possible 
is the opposite orientation of magnetic loops in the northern and southern 
hemispheres, in contrast with most previous 2D simulations 
which were limited to only one hemisphere. 
Strong mixing between the 
two hemispheres, discovered already in \citet{camus-09}, reduces the 
characteristic length scale of magnetic inhomogeneities and speeds up the 
magnetic dissipation.  The displacement of magnetic loops near the polar 
axis via the kink instability is an additional way of facilitating 
magnetic dissipation in 3D models. This agrees with the observed lower 
level of magnetic energy inside PWN in these models 
(see Figure~\ref{fig:mdiss}).         
          
The data presented in Figure~\ref{fig:2Dvs3D} show that the axial 
compression remains a feature of our 3D-solution near the termination 
shock and that the ``tooth-paste'' effect of hoop-stress due to   
freshly-injected azimuthal magnetic field of the pulsar wind is still 
capable of driving polar jets. But compared to the 2D models, the region 
of operation of this jet-production mechanism is much more compact.

\section{Discussion}

The results of our simulations provide strong support to Begelman's 
hypothesis that the essence of the sigma-problem of PWN relies in the inadequacy 
of using over-simplified models. He argued that these models were unstable 
to kink modes and that randomization of magnetic field, resulting from 
this instability, would lead to a termination shock of size similar to that of 
weakly-magnetized models, even if the actual magnetization was still very 
high. Our results confirm, and augment this by showing that the magnetization 
may actually be significantly reduced via onset of magnetic dissipation 
accompanying randomization of magnetic field. The observed properties 
of PWN may mimic those of models with weakly magnetized winds not because 
magnetic stress of randomized magnetic field effectively reduces 
to pressure but because the magnetic dissipation renders the PWN 
gas-pressure-dominated (high-$\beta$ plasma).  
In fact, high-magnetization 
of PWN plasma seems to be ruled out by the observations of synchrotron and 
inverse-Compton emission of the Crab nebula, which show that the magnetic field 
is energetically sub-dominant to the population of relativistic electrons 
by a factor of $\sim 30$ \citep{hillas-98}.               
    
Although the dissipation in the simulations ultimately is of numerical origin and
occurs at the cell-size scale, the extreme resolution employed in a fully conservative, grid-adaptive computation can mimic processes creating such fine
magnetic structures and hence the energy conversion has to be taken
seriously. Moreover, the observed energy content of the Crab nebula
does suggest efficient magnetic dissipation on a time scale which is
significantly shorter than the age of the nebula 
but much longer compared to its Alfv\'en crossing time \citep{ssk-mdiss-12}.

The dynamics of the inner region of our 3D numerical models is still 
strongly influenced by the regular magnetic field of freshly-injected
plasma. The hoop stress of this field is still capable of producing 
noticeable axial compression close to the termination shock and driving polar 
outflows, required to explain the Crab jet, and jets of other PWN. However, 
these are much more moderate than in 2D models. It appears that the 
strong relativistic jets emerging in previous 2D simulations of magnetar bubbles, 
with application to Gamma Ray Bursts, may well be artifacts of the imposed 
axial symmetry.             

\section{Acknowledgments}
SSK and OP are supported by STFC under the standard grant ST/I001816/1. 
SSK acknowledges support by the Russian Ministry of Education and Research under the
state contract 14.B37.21.0915 for Federal Target-Oriented Program.
RK  acknowledges FWO-Vlaanderen, grant G.0238.12, and BOF F+ financing related to EC FP7/2007-2013 grant agreement SWIFF (no.263340).

\bibliographystyle{mn2e}
\bibliography{pwn,mypapers,lyubarsky,lyutikov}
\end{document}